\documentclass[a4paper,11pt]{article}

\usepackage{color}

\usepackage[]{array}
\usepackage[]{mathrsfs}
\usepackage[]{tensor}
\usepackage[symbol]{footmisc}

\newcommand{\str}{\textrm{str}}
\newcommand{\qqd}{\ , \quad}

\newcommand{\ctg}{\textrm{ctg}}
\newcommand{\be}{\begin{eqnarray}}
\newcommand{\ee}{\end{eqnarray}}
\newcommand{\0}{\nonumber}
\newcommand{\Gam}{\mathbf{\Gamma}}
\newcommand{\R}{\mathbf{R}}
\newcommand{\N}{N}
\newcommand{\NGAUG}{S}
\newcommand{\A}{\mathbf{A}}

\newcommand{\F}{\mathbf{F}}

\newcommand{\K}{\mathbf{K}}

\newcommand{\LL}{\mathbf{L}}

\newcommand{\CS}{\mathbf{\Upsilon}}

\newcommand{\1}{I}
\newcommand{\refb}[1]{(\ref{#1})}
\newcommand{\tr}{\rm tr}
\begin{document}
\begin{flushright}
{SISSA 23/2011/EP \\ ZTF-11-05}\\
{hep-th/1105.4792}
\end{flushright}
\vskip 1cm
\begin{center}
{\LARGE{\bf Gravitational Chern-Simons Lagrangian terms and spherically 
symmetric spacetimes}}
\vskip 1cm

{\Large L.~Bonora$^{a,b}$, M.~Cvitan$^c$, P.~Dominis Prester$^d$, S.~Pallua$^c$, I.~Smoli\'c$^{a,c}$}\\
{}~\\
\quad \\
{\em ~$~^{a}$International School for Advanced Studies (SISSA/ISAS),}\\
{\em Via Bonomea 265, 34136 Trieste, Italy}
{}~\\
\quad \\
{\em ~$~^{b}$INFN, Sezione di Trieste, Italy}
{}~\\
\quad \\
{\em ~$~^{c}$Theoretical Physics Department, Faculty of Science,}\\
{\em University of Zagreb, p.p.~331, HR-10002 Zagreb, Croatia}
 {}~\\
\quad \\
{\em ~$~^{d}$ Department of Physics, University of Rijeka,}\\
{\em  Omladinska 14, HR 51000 Rijeka, Croatia}\\
\vskip 1cm
Email: bonora@sissa.it, mcvitan@phy.hr, pprester@phy.uniri.hr, pallua@phy.hr, ismolic@phy.hr

\end{center}

\vskip 3cm {\bf Abstract.}  We show that for general spherically symmetric configurations, contributions of {broad class of} gravitational and mixed gauge-gravitational Chern-Simons terms to the equations of motion vanish identically in  $D>3$ dimensions. This implies that such terms in the action do not affect Birkhoff's theorem or any previously known spherically symmetric solutions. Furthermore, we investigate the thermodynamical properties using the procedure described in an accompanying paper. We find that in $D > 3$ static spherically symmetric case Chern-Simons terms do not contribute to the entropy either. Moreover, if one requires only for the metric tensor to be spherically symmetric, letting other fields unrestricted, the results extend almost completely, with only one possible exception --- Chern-Simons Lagrangian terms in which the gravitational part is just the $n=2$ irreducible gravitational Chern-Simons term.
\vskip 1cm 


\vfill\eject

\section{Introduction}%

In a companion paper, \cite{BCDPS}, referred to as \1, we have started the study of gravitational Chern-Simons terms in higher dimensions. The motivation comes from the observation that, while the 2+1 dimensional case can count on a considerable number of analyses \cite{DJT1,DJT2,Solodukhin:2005ah,Perez:2010hk,Kraus:2005zm,Park:2006gt,Miskovic:2009kr}, little is known about higher dimensional gravitational Chern-Simons theories (we will specify shortly what we mean with this terminology). In \1 we started searching for systematic answers to the questions raised by the presence of these terms in higher dimensions.
One of the problems they raise is how their addition modifies black hole solutions and associated charges. Another important problem is how to compute the black hole entropy in their presence. In this paper, relying in particular on the results of \1, we address these issues for spherically symmetric configurations.
 
Let us briefly review the above mentioned problems by introducing a few definitions and basic properties. We wish to investigate the properties of gravitational actions extended with Chern-Simons terms,
\be \label{lagrgen}
\LL = \LL_{\mathrm{cov}} + \LL_{\mathrm{CS}}
\ee
By $\LL_{\mathrm{cov}}$ we denote some generic manifestly diffeomorphism covariant gravitational Lagrangian $D$-form in $D$ dimensions, while $\LL_{\mathrm{CS}}$ contain Chern-Simons terms, which are, on the contrary, not manifestly covariant.

A general gravitational Chern-Simons (CS) term, in $D = 2n - 1$ dimensions, has the form
\be\label{LCS}
\CS_n(\Gam) = n \int_0^1 dt \ P_n (\Gam, \R_t, \dots, \R_t)
\ee
where $\R_t= t d\Gam + t^2 \Gam\Gam$, $\Gam$ is the Levi--Civita connection and $P_n$ denotes an invariant symmetric polynomial of the appropriate Lie algebra, which, for purely gravitational CS terms, is the Lie algebra of the $SO(1,D-1)$ group (in this case $P_n$ are symmetrized traces). In general the polynomial $P_n$ may be irreducible or reducible. It is important to recall that for $n=2k-1$, that is, in 
$D = 4k - 3$ dimensions, {\it irreducible} invariant symmetric polynomial of the Lie algebra of $SO(1,D-1)$ identically vanish, $P_{2k-1}= 0$.
So, purely gravitational irreducible CS terms can appear only in $D = 4k- 1$ dimensions. In this paper we will consider both reducible and irreducible $P_n$'s.\footnote{Our notation is mainly as in \cite{Bertl}.}

We will also consider action terms where a gravitational CS term is multiplied by an invariant polynomial made of one or more gauge field strengths, so as to fill up a $D$-form. The simplest of such reducible action terms have the form
\be\label{mixCS}
\LL_{1,\mathrm{mix}} = \CS_m(\Gam) P_k(\F) \quad \textrm{or} \quad 
\LL_{2,\mathrm{mix}} = \CS_k(\A) P_m(\R)
\ee
where $\F$ represents the generic curvature of a gauge connection $\A$ and $P_k$ is an invariant polynomial of order $k$, such that $D= 2m+2k-1$. $\A$ can  be either a non-Abelian or an Abelian gauge connection, or even an RR field (in the latter case the relation between form orders and spacetime dimension may be different from the just mentioned one). We will refer to terms like (\ref{mixCS}) as {\it mixed Lagrangian} terms.

To summarize and state our problem with precision, in this paper we will consider a broad class of CS terms
\be \label{LCSgen}
\LL_{\mathrm{CS}} = \sum_i \LL_{\mathrm{gCS}}^{(i)} + \sum_j \LL_{\mathrm{aCS}}^{(j)}
\ee
where
\be
\LL_{\mathrm{gCS}}^{(i)} &=& \CS_{n_{i1}}(\Gam) \, P_{n_{i2}}(\R) \, P_{n_{i3}}(\R) \cdots \,
  P_{r_{i1}}(\F_1) \, P_{r_{i2}}(\F_2) \cdots
\label{mixgCS} \\
\LL_{\mathrm{aCS}}^{(j)} &=& \CS_{m_j}(\A_{j}) \, P_{n_{j1}}(\R) \, P_{n_{j2}}(\R) \cdots \,
  P_{r_{j1}}(\F_1) \, P_{r_{j2}}(\F_2) \cdots
\label{mixaCS}
\ee
and $\Gam$ is the symmetric Levi-Civita connection. It is assumed in (\ref{mixaCS}) that there is at least one $P_n(\R)$ term present. Also, it is understood in (\ref{mixgCS}) and (\ref{mixaCS}) that indices $n$ 
are even integers, otherwise, as explained before, the terms vanish identically (up to total derivatives). Terms present in (\ref{LCSgen}), which are of gravitational and mixed gauge-gravitational type, for brevity we shall sometimes call gravitational CS Lagrangian terms.

For completeness, we note that purely gauge Chern-Simons Lagrangian terms, which are generally of the form
\be \label{gaugeCS}
\LL_{\mathrm{CS}}^{(j)} = \CS_{m_j}(\A_{j}) \, P_{r_{j1}}(\F_1) \, P_{r_{j2}}(\F_2) \cdots \ ,
\ee
due to manifest diff-covariance are, if present, contained in the $\LL_{\mathrm{cov}}$ part of Lagrangian 
(\ref{lagrgen}). The CS terms (\ref{mixaCS}) too are manifestly diff-covariant, but due to their common origin with (\ref{mixgCS}) it is useful to treat them jointly.

In \1 we have analyzed the consequences of adding gravitational or mixed CS terms in gravity theory,
as far as black hole entropy is concerned. Extending the covariant phase space formalism of
\cite{IyerWald}, according to \cite{Tach}, we have computed in detail the CS induced modifications to Wald entropy formula. We have analyzed the covariance of the ensuing entropy formula and concluded that, although it looks superficially non-covariant, it can be cast in a covariant form.  

In the present paper we continue the analysis of the consequences of adding CS terms 
(\ref{LCSgen}){-(\ref{mixaCS})} to gravitational actions, by considering the specific case of spherically symmetric metrics. We will show that solutions with spherically symmetric metric are generally not modified in $D>3$ dimensions, with the only possible exceptions when the following two conditions are simultaneously present: (i) a gravitational CS Lagrangian term, which is a wedge-product of the irreducible gravitational $n=2$ CS term and a purely gauge factor, is present, (ii) this gauge factor (which is by definition gauge invariant) is not spherically symmetric for the solution in question. If both conditions are fulfilled then this ``exceptional'' gravitational CS Lagrangian term may possibly modify such solution of $\LL_{\mathrm{cov}}$.

Moreover, relying on the results of \1, we will show that the black hole entropy is not modified either, with the same exception understood.   

The paper is organized as follows. In Sec.\ \ref{sec:eom} we write down the contributions to the equations of motion
of CS Lagrangian terms present in (\ref{LCSgen}). In Sec.\ \ref{sec:sssol} we analyze these contributions in the case of spherically symmetric metric, where the central results are stated in Theorems 1 and 2. In Sec.\ 
\ref{sec:entropy} we analyze contributions to black hole entropy, the main result being stated in Theorem 3. In Sec.\ \ref{sec:concl} we summarize our findings. In Appendices some technical aspects of the calculations are presented in more detail.

\section{Equations of motion}
\label{sec:eom}

Adding gravitational CS terms in the Lagrangian brings about additional terms in the equations of motion. It was shown in \cite{Solodukhin:2005ns} that the equation for the metric tensor $g_{\alpha\beta}$ acquires an additional term 
$C^{\alpha\beta}$ of the form
\begin{equation} \label{CSeom}
C^{\alpha\beta} = \nabla_{\!\rho} \, S^{(\alpha\beta)\rho}
\end{equation}
where tensor the $S^{(\alpha\beta)\rho}$, whose exact form depends on the explicit form of the CS Lagrangian terms,
is antisymmetric in the last two indices. 
Moreover the tensor $C^{\alpha\beta}$ is traceless and covariantly conserved
\be
\tensor{C}{^\alpha_\alpha} = 0 \qqd \nabla_{\!\alpha} \, C^{\alpha\beta} = \nabla_{\!\alpha} \nabla_{\!\rho} \, S^{(\alpha\beta)\rho} = 0
\ee
and $S$ satisfies the following properties
\be
S^{\alpha\beta\rho} = S^{\alpha[\beta\rho]} \qqd \tensor{S}{^\alpha^\beta_\alpha} = 0 \qqd \nabla_{\!\alpha} \, S^{\alpha\beta\rho} = 0
\ee
These follow from the symmetries of the Riemann tensor, Bianchi identities
and the form of $S$.
It was also shown in \cite{Solodukhin:2005ns} that the tensor $C^{\alpha\beta}$ can be viewed as a generalization of the Cotton tensor.

Mixed CS Lagrangian terms contribute also to equations of motion for gauge fields participating in these terms. 
What follows is an overview of contributions to the equation of motion of different types of gravitational CS terms.

We consider a mixed Lagrangian with one gravitational CS term, $\N-1$ gravitational terms and $\NGAUG$ gauge terms,
\be
\LL = \CS_{n_1}(\Gam) \, P_{n_2}(\R) \cdots P_{n_{\N}}(\R) \,\, P_{r_1}(\F_1) \cdots P_{r_{\NGAUG}}(\F_\NGAUG)
\ee
where each gauge field strength $F_i$ is a $p_i$-form. Using partial integration and discarding boundary terms, 
the variation of this Lagrangian can be put in the following form
\be \label{genEOMs}
\delta \LL &=& \sum_{i=1}^\N P_{n_1}(\R)  \cdots \delta\CS_{n_i}(\Gam) \cdots P_{n_{\N}}(\R) \, P_{r_1}(\F_1) \cdots P_{r_\NGAUG}(\F_\NGAUG) + \\
&+& \sum_{j=1}^{\NGAUG} (-1)^{1+\gamma_j} \, P_{n_1}(\R) \cdots P_{n_\N}(\R)  \, P_{r_1}(\F_1) \cdots  \delta\CS_{r_j}(\A_j)  \cdots P_{r_{\NGAUG}}(\F_{\NGAUG})\0
\ee
Here $P_{n_1}(\R)  \cdots \delta\CS_{n_i}(\Gam) \cdots P_{n_{\N}}(\R)$ denotes the wedge product 
$P_{n_1}(\R) \cdots P_{n_{\N}}(\R)$ in which $P_{n_{i}}(\R)$ is replaced by $\delta\CS_{n_i}(\Gam)$.
The order of the form $P_{r_1}(\F_1) \cdots \, P_{r_{j-1}}(\F_{j-1})$ is denoted by $\gamma_j$ and is equal to $\gamma_j = \sum_{l=1}^{j-1} p_l r_l$.

The first line of \refb{genEOMs} determines the equation of motion \refb{CSeom} for the metric and
the second line the equations of motion for the gauge fields. 
Now we write the equation of motion for the metric in more detail.
To write it in compact form we first 
introduce the $2(n-1)$-form $\left(\mathbf{K}_{(n)}\right)^{\alpha\beta} \equiv \left(\R^{n-1}\right)^{\alpha\beta}$, whose components are
\be
K_{(n)}^{\alpha\beta}{}_{\mu_1 \cdots \mu_{2n-2}} = \frac{(2n-2)!}{2^{n-1}} \, \tensor{R}{^\alpha_{\sigma_1}_[_{\mu_1}_{\mu_2}} \, \tensor{R}{^{\sigma_1}_{| \sigma_2 |}_{\mu_3}_{\mu_4}} \cdots \tensor{R}{^{\sigma_{n-2}}^{\beta}_{\mu_{2n-3}}_{\mu_{2n-2}]}}
\ee
Using the symmetries of the Riemann tensor, and the fact that $n$ is even, it can be easily shown that this tensor is antisymmetric in its upper indices.
It is also convenient to introduce the $(2n-1)$-form $\bar{\K}{}^{\alpha\beta\gamma}_{(n)}$ whose components are
\be
\bar{K}_{(n)}^{\alpha\beta\gamma}{}_{{\mu_1} \cdots {\mu_{2n-1}}} = (2n-1)K_{(n)}^{\alpha\beta}{}_{[\mu_1 \cdots {\mu_{2n-2}} } \delta^\gamma_{\mu_{2n-1}]}
\ee
The generic variation of $\CS_n(\Gam)$ (see \1) can now be written as 
\be\label{deltaLCS2}
\delta \CS_n(\Gam) = n \, P_n (\delta \Gam, \R^{n-1}) + d(\ldots) = n \, \bar{\K}_{(n)}^{\alpha}{}_\beta{}^\gamma \delta \Gamma^\beta{}_{\alpha \gamma} + d(\ldots)
\ee
The part of variation which is exact does not affect the equation of motion, so we will dispense
from writing it down explicitly%
\footnote{The interested reader can find these boundary terms, 
which contribute to the symplectic potential, written down explicitly in \cite{BCDPS}.}.

Now, using
\be
\delta \Gamma^\lambda_{\alpha \mu} = \frac{1}{2} \, g^{\lambda \beta} \left( \nabla_{\!\alpha} \, \delta g_{\beta \mu} + \nabla_{\!\mu} \, \delta g_{\beta \alpha} - \nabla_{\!\beta} \, \delta g_{\alpha \mu} \right)
\ee
together with antisymmetry of $\bar{\K}_{(n)}^{\alpha\beta\gamma}$ in $\alpha$ and $\beta$, we obtain (up to boundary terms)
\be
\delta \CS_n(\Gam) =   n \,
\bar{\K}_{(n)}^{\rho\alpha\beta} \, \nabla_{\!\rho} \delta g_{\alpha\beta} 
\ee
Comparing the first line of \refb{genEOMs} (after the partial integration) to 
$\nabla_\rho S^{\alpha\beta\rho} \delta g_{\alpha \beta} {\mathbf{\epsilon}}$ we obtain the tensor $S$
\be \label{genS}
S^{\alpha\beta\rho} = (-)^{s+1}\sum_{i=1}^\N *\left( n_i \,
P_{n_1}(\R)  \cdots \bar{\K}_{(n_i)}^{\rho\alpha\beta} \cdots P_{n_{\N}}(\R) \, P_{r_1}(\F_1) \cdots P_{r_\NGAUG}(\F_\NGAUG)
\right)
\ee
where $*$ denotes the Hodge dual\footnote{$(*A)^{a_{p+1}\ldots a_D} = \frac{1}{p!} A_{a_1\ldots a_p} \epsilon^{a_1 \ldots a_D}$}. 
Here $s$ denotes the number of minuses in the metric signature. Note that the form in the brackets on the right hand side is a $D$-form.

In the case $\N=1$ and $\NGAUG=0$, i.e.\ irreducible pure gravity CS, the last expression reduces to:
\be
S^{\alpha\beta\rho} = (-)^{s+1}* \left( n \bar{\K}_{(n)}^{\rho\alpha\beta} \right)
=  -\frac{(-)^s n}{(2n-2)!} \, \epsilon^{\mu_1 \ldots \mu_{2n-2}\beta} K_{(n)}^{\rho\alpha}{}_{\mu_1 \ldots {\mu_{2n-2}} } 
\ee
The contribution to the equation of motion is then 
\be \label{Cirred}
C^{\alpha\beta} = \frac{(-1)^s n}{(2n-2)!} \ \epsilon^{\mu_1 \cdots \mu_{2n-2} (\alpha} \, \nabla_{\!\rho} K_{(n)}^{\beta)\rho}{}_{\mu_1 \ldots {\mu_{2n-2}}}
\ee

As a simple example, let us apply (\ref{Cirred}) to the pure gravity irreducible CS Lagrangian term in 
$D=7$ dimensions.  Its explicit contribution to the action is
\be
S_7 &=& \int \CS_4 
 = \int \str( \R^3 \Gam - \frac{3}{5} \R^2 \Gam^3 + \frac{1}{5} \R   \Gam^5 - \frac{1}{35}\Gam^7)
\label{s7}  \\
&=& \int \tr( \R^3 \Gam
-\frac{2}{5} \R^2 \Gam^3
-\frac{1}{5} \R   \Gam^2  \R  \Gam
+\frac{1}{5} \R   \Gam^5
-\frac{1}{35}\Gam^7) \0
\ee
and to the equations of motion is
\be \label{C7}
C^{\alpha\beta} = \frac{(-1)^s }{2} \ \epsilon^{\mu_1 \cdots \mu_{6} (\alpha} \, \nabla_{\!\rho}  \, \left( \tensor{R}{^{\beta)}_{\sigma_1}_{\mu_1}_{\mu_2}} \, \tensor{R}{^{\sigma_1}_{ \sigma_2 }_{\mu_3}_{\mu_4}} \tensor{R}{^{\sigma_{2}}^{\rho}_{\mu_{5}}_{\mu_{6}}} \right) 
\ee

\section{Spherically symmetric solutions}
\label{sec:sssol}

The most general spherically symmetric $D$-dimensional metric can be written (see Appendix B of \cite{HawkingEllis}) in the following form,
\be\label{sphsym}
ds^2 = -f(t,r) \, dt^2 + \frac{dr^2}{g(t,r)} + h(t,r) \, d\Omega^2_{D-2}
\ee
We are using here coordinates
\be \label{gensphcoord}
x^0 = t \qqd x^1 = r \qqd x^i = \theta^i \quad (i = 2, 3, \dots , D-1)
\ee
where $\theta^i$ are angular coordinates on spheres defined by $t,r=\textrm{constant}$. Angular coordinates are such that
$$0 \le \theta^i < \pi \quad \textrm{for} \quad i = 2, \dots, D-2 \quad \textrm{and} \quad 0 \le \theta^{D-1} < 2\pi$$
(this last coordinate $\theta^{D-1}$ is more frequently denoted by $\phi$). In the rest of the paper we will use indices $i$, $j$ and $k$ for angular coordinates ($i,j,k = 2, \dots, D-1$). Introducing the auxiliary function
\be
\Pi(k) = \left\{ \begin{array}{ccl}  1 & , & k = 2 \\ & & \\ \prod_{m=2}^{k-1} \, \sin^{2} \theta^m  & , & k \ge 3 \end{array} \right.
\ee
we can write the metric components in a simple way as
\be
g_{00} = - f(t,r) \qqd g_{11} = \frac{1}{g(t,r)} \qqd g_{ii} = h(t,r) \, \Pi(i) \qqd i = 2, \dots , D-1
\ee
In the argument that follows, the crucial piece of information consists in identifying the nonvanishing components of the Riemann tensor for this type of metric.\\

\noindent
\textbf{Lemma 1.} The only non-vanishing components of the Riemann tensor for a spherically symmetric metric in coordinates \refb{gensphcoord} are, up to symmetries of their indices, of the form $R_{\mu\nu\mu\nu}$ and $R_{0i1i}$. Since the metric is diagonal, this remains valid even if some of the indices are raised.\\

\noindent
From this property immediately it follows that
\be \label{rrr0}
\R^3 = 0  \quad\quad \quad  \tr (\R^2) = 0
\ee
The proof is by direct calculation of the Riemann tensor components, and is presented in Appendix B. Using Lemma 1 we can look at the product of Riemann tensors inside the tensor 
$C^{\mu\nu}$. A consequence is the following theorem:\\

\noindent
\textbf{Theorem 1.} \ The contribution of the gravitational and mixed gauge-gravitational Chern-Simons Lagrangian terms (\ref{LCSgen}){-(\ref{mixaCS})} to the equations of motion vanishes identically for any configuration with spherically symmetric metric in $D > 3$ dimensions, unless the gravitational contribution is the sole $\CS_2(\Gam)$ factor wedge-multiplied by the spherically asymmetric gauge factor (we refer to such terms in Lagrangian as \emph{exceptional} terms).\\

\emph{Proof}: First notice that in all cases, except in those in which $\CS_2(\Gam)$ is present as the only gravitational factor, 
all contributions to the equations of motion of the CS terms contain as a factor either 
$\R^3$ or $\tr (\R^2)$, which, by Lemma 1, vanish for spherically symmetric metrics (\ref{sphsym}). In $D>3$ this leaves only the following gravitational CS Lagrangian terms as potentially nontrivial
\begin{equation} \label{gcs2mix}
\CS_2(\Gam) \, P_{r_1}(\F_1) \cdots P_{r_\NGAUG}(\F_\NGAUG) \ ,
\end{equation}
where $F_i$ are some $p_i$-form gauge field strengths. The total contribution of such terms to the Lagrangian is obviously of the form
\be \label{gcs2tot}
\CS_2(\Gam) \, G(\F) \;\;,
\ee
where $G(\F)$ is a gauge invariant $(D-3)$-form. From (\ref{genEOMs}) it follows that (\ref{gcs2tot}) contributes only to the equation of motion for the metric $g_{ab}$ (due to the $\str(\R^2)=0$ factor appearing in the equations for gauge fields). The tensor $S^{\alpha\beta\rho}$ is 
\be \label{Sspher}
S^{\alpha\beta\rho} &=& (-)^{s+1} *\left( 2 \bar{\K}_{(2)}^{\rho\alpha\beta}  G(\F) \right) \0\\
&=& (-)^{s+1} \, \frac{2\cdot 3}{D!} \, \epsilon^{\mu_1\cdots\mu_{D-1}\beta}
 R^{\rho\alpha}{}_{\mu_1\mu_2} \, G(\F)_{\mu_{3}\cdots\mu_{D-1}},
\ee
which follows from \refb{genS}.
Following the statement in Theorem 1, we now restrict ourselves to configurations of gauge fields for which $G(\F)$ is spherically symmetric. 
Spherically symmetric forms correspond to one, or to the linear combination of two, of the following cases:
\begin{itemize}
\item[(a)] $a(t,r) \, dt + b(t,r) \, dr \quad$ (1-form)
\item[(b)] $a(t,r) \, dt \wedge dr \quad$ (2-form)
\item[(c)] $a'(t,r) \, \tilde{\epsilon}_{D-2} \quad$ ($(D-2)$-form)
\item[(d)] $a'(t,r) \, dt \wedge \tilde{\epsilon}_{D-2} + b'(t,r) \, dr \wedge \tilde{\epsilon}_{D-2}
 \quad$ ($(D-1)$-form)
\item[(e)] $a'(t,r) \, dt \wedge dr \wedge \tilde{\epsilon}_{D-2} \quad$ ($D$-form)
\end{itemize}
where $\tilde{\epsilon}_{D-2}$ is the volume form of $S^{D-2}$ sphere. Since $G(\F)$ is $(D-3)$-form, the possibilities (c)-(e) are trivially excluded. 
Thus we are left with the first two, mutually excluding, possibilities (a) and (b). 
For (a), which exists only in $D=4$ dimensions, and where $G$ is a derivative of some scalar field, it was shown in the literature \cite{Grumiller:2007rv} that the spherical symmetry of the metric forces 
$C^{\alpha\beta}=0$, i.e., there is no contribution to the equations of motion.\footnote{Results from 
\cite{Cantcheff:2008qn} suggest that this may not be valid in Einstein-Cartan first-order approach to gravity. However, in this paper we stick to standard general relativistic formulation of gravity with torsionless connection.} For (b), which exists only in $D=5$ dimensions, by using $G(\F) \propto dt \wedge dr$ and expressions for components of the Riemann tensor (\ref{Rspher}), we obtain that the only nonvanishing components of the $S^{\alpha\beta\rho}$ tensor have the form $S^{ijk} \propto \epsilon^{ijk01}$. We see that $S^{\alpha\beta\rho}$ is antisymmetric in the first two indices, so $S^{(\alpha\beta)\rho} = 0$, which by (\ref{CSeom}) gives $C^{\alpha\beta}=0$. Thus we have obtained that terms of the type (\ref{gcs2mix}) will not contribute either to the equations of motion if, in addition to the metric tensor, the total gauge factor $G(\F)$ is spherically symmetric. This completes the proof 
of Theorem 1.\\

Theorem 1 extends the recent result from \cite{LuPang} and states that addition of any combination of gravitational and mixed gauge-gravitational Chern-Simons terms 
{(\ref{LCSgen})-(\ref{mixaCS})} to any Lagrangian leaves all spherically symmetric solutions unchanged in $D>3$ dimensions.\footnote{By spherically symmetric configurations we mean configurations in which metric and gauge-invariant tensors, such as $G(\F)$, are spherical 
symmetry invariants.} It also shows that this is still valid in much broader circumstances. Even if we require only for the metric to be spherically symmetric, and allow other fields to be spherically asymmetric, Theorem 1 states that solutions can be affected only if two conditions are met: (i) CS terms of the type 
(\ref{gcs2tot}) are added to Lagrangian, (ii) the total gauge part (of such terms collected together) $G(\F)$ evaluated on the solution is \emph{not} a spherically symmetric $(D-3)$-form. We postpone further discussion on these ``exceptional'' cases to Sec. \ref{sec:concl}.

For completeness, we mention that purely gauge CS Lagrangian terms (\ref{gaugeCS}) cannot be 
included in the general statement of Theorem 1. It was shown in the literature on some explicit 
examples that such CS Lagrangian terms can affect spherically symmetric solutions 
\cite{Brihaye:2009cc,Brihaye:2010wp,Brihaye:2011nr}. However, we remind the reader that purely 
gauge CS terms are manifestly diff-covariant and so do not fall in the class of CS terms we are 
interested in this paper.\\

Although the $D=3$ case is not meant to be explicitly addressed in this paper, let us
summarize what can be said about it on the basis of the existing literature.
In $D=3$ dimensions Theorem 1 is in general not valid. Here one has the $n=2$ irreducible gravitational Chern-Simons term with contribution to equations of motion proportional to the Cotton-York tensor,
\be\label{Cotton}
C^{\mu\nu} = \epsilon^{\mu\alpha\beta} \, \nabla_{\!\alpha} \left( \tensor{R}{^\nu_\beta} - \frac{1}{4} \, \tensor{\delta}{^\nu_\beta} R \right)
\ee
which does not vanish for a general spherically symmetric metric (\ref{sphsym}). For example, it was shown in \cite{Garcia} that already for a static spherically symmetric metric
\be
ds^2 = -f(r) dt^2 + \frac{dr^2}{f(r)} + r^2 d\phi^2
\ee
one obtains
\be
C^{t\phi} = \frac{f'''}{4\sqrt{f}},
\ee
which is generally non-vanishing.

However, it is known that the Cotton tensor vanishes for several important classes of 3D metrics, so in these cases the 3D gravitational CS Lagrangian term is effectively irrelevant as far as the equations of motion are concerned. For example, the Cotton tensor vanishes for any 3D Einstein metric\footnote{$g_{ab}$ is an \emph{Einstein metric} if the Ricci tensor is proportional to the metric itself, $R_{ab} = k g_{ab}$ for some \emph{constant} $k$. It is true in any dimension $D \ge 3$ that $g_{ab}$ is an Einstein metric if and only if it is a solution to the vacuum Einstein field equations with cosmological constant $\Lambda = (D-2)k/2$.}. In particular, the Cotton tensor vanishes for the famous BTZ black hole solution \cite{Kaloper:1993kj}. The generalized Cotton tensor is still conformally invariant \cite{Solodukhin:2005ns}, but the Einstein metrics in $D > 3$ are not necessary maximally symmetric.\\

Birkhoff's theorem in the presence of gravitational CS term in $D=3$ was examined by Cavagli\`a 
\cite{Cavaglia}, and in particular class of $D=4$ theories with mixed-type gravitational CS Lagrangian term in \cite{Yunes:2007ss}. Using our Theorem 1 we generalize these results to a wide range of higher dimensional cases:\\

\noindent
\textbf{Theorem 2.} (\textsf{Chern-Simons-Birkhoff}) \ Assuming that we have a theory in $D>3$ dimensional spacetime (described by the Lagrangian $\LL_G$) in which all spherically symmetric solutions are necessarily static, this property remains valid even if we include additional gravitational (pure or mixed) Chern-Simons Lagrangian terms $\LL_{CS}$ (\ref{LCSgen}){-(\ref{mixaCS})}.\\

For example, Birkhoff's theorem is known to be valid for the Lovelock--type gravitational Lagrangians in any dimension (see \cite{Zegers,Deser:2005gr}). Some other examples can be found in \cite{Oliva:2011xu}.\\

\section{Entropy of static spherically symmetric black holes}
\label{sec:entropy}

In the previous section we have shown that, generically, gravitational CS Lagrangian terms do not affect spherically symmetric solutions in $D>3$. Here we will show that they do not affect the entropy of static spherically symmetric black holes. By applying staticity to (\ref{sphsym}) one obtains that any static spherically symmetric metric can be written in the form
\be\label{staticm}
ds^2 = -f(r) dt^2 + \frac{dr^2}{g(r)} + h(r) d\Omega_{D-2}^2
\ee
which can be used for static spherically symmetric black holes outside the horizon.

In the theories we consider the Lagrangian (\ref{lagrgen}) can be written in the following forms
\begin{equation} \label{Lgen}
\LL = \LL_{\mathrm{cov}} + \LL_{\mathrm{CS}}
 = \LL_{\mathrm{cov}} + \LL_{\mathrm{aCS}} + \LL_{\mathrm{gCS}}
 = \LL_{\mathrm{diff-cov}} + \LL_{\mathrm{gCS}}
\end{equation}
where the part $\LL_{\mathrm{diff-cov}}$ contains all manifestly diff-covariant terms
(which includes also $\LL_{\mathrm{aCS}}$ part), while the second piece
contains all non-manifestly diff-covariant terms. In the class of theories we consider in this paper, $\LL_{\mathrm{gCS}}$ is made of terms as in (\ref{mixgCS}). In such theories the entropy assigned to the black hole solutions can correspondingly be split in several ways
\begin{equation} \label{entgen}
S_{\mathrm{bh}} = S_{\mathrm{cov}} + S_{\mathrm{CS}}
 = S_{\mathrm{cov}} + S_{\mathrm{aCS}} + S_{\mathrm{gCS}}
 = S_{\mathrm{diff-cov}} + S_{\mathrm{gCS}}
\end{equation}
The piece $S_{\mathrm{diff-cov}}$ can be obtained from $\LL_{\mathrm{diff-cov}}$, and is given by the general Wald formula \cite{Wald1,JKM,IyerWald}
\be \label{Swald}
S_{\mathrm{diff-cov}} = S_W = 2\pi \int_{\mathcal{B}} \epsilon^a{}_b
 \left( \frac{\delta \LL_{\mathrm{diff-cov}}}{\delta \tensor{R}{^a_b_\mu_\nu}} \right)_{\mu\nu\rho_1\cdots\rho_{D-2}}
\ee
The second piece in (\ref{entgen}), $S_{\mathrm{gCS}}$, can be obtained from 
$\LL_{\mathrm{CS}}$ by the generalization of Wald's procedure to non-manifestly 
diff-covariant Lagrangians, as described in \cite{Tach}. In \cite{BCDPS} it was 
shown that for the gravitational CS Lagrangian terms (\ref{mixgCS}) there is a general expression also for this part of the entropy 
\be \label{genwaldent}
S_{\mathrm{gCS}} = 2\pi \int_{\mathcal{B}} \tensor{\epsilon}{^a_b}\left( \frac{\delta \LL_{\mathrm{gCS}}}{\delta \tensor{\R}{_t^a_b_\mu_\nu}} \right)_{\mu\nu\rho_1\cdots\rho_{D-2}}
\ee
which is similar in form to Wald formula (\ref{Swald}), the only difference being that one takes the variation with respect to $\R_t$ instead of $\R$.\footnote{Here it is understood that the variation with respect to $\R_t$ acts inside the 
$t$ integral present in the Lagrangian.
The Lagrangians used here are of the form $\LL_{\mathrm{gCS}}^{(i)}$ from (\ref{LCSgen}), which means that there will always be exactly one $\CS$ factor, and consequently exactly one $t$ integral.}

In both terms (\ref{Swald}) and (\ref{genwaldent}) one inserts the black hole solution to the equations of motion obtained from the complete Lagrangian (\ref{Lgen}), and integrates over the $(D-2)$-dimensional horizon cross-section (bifurcation surface) with binormal 
$\epsilon_{ab}$, normalized by $\epsilon_{ab}\epsilon^{ab} = -2$.\\

\noindent
\textbf{Theorem 3.} \ Gravitational CS Lagrangian terms {(\ref{LCSgen})-(\ref{mixaCS})} do not affect the entropy of static spherically symmetric black holes\footnote{Static spherically symmetric black holes are
defined here as being characterized by a static and spherically symmetric metric tensor(\ref{staticm}), with no conditions on the other fields.} in $D > 3$ dimensions, apart from the exceptional cases mentioned in Theorem 1 (which have spherically asymmetric gauge field configurations).\\

\emph{Proof}: First we note that Theorem 1 guarantees (up to the exceptional cases mentioned below) that gravitational CS Lagrangian terms {(\ref{LCSgen})-(\ref{mixaCS})} do not change static spherically symmetric black hole solutions (obtained from $\LL=\LL_{\mathrm{cov}}$), which means that they do not change the $S_{\mathrm{cov}}$ part of the black hole entropy. The only possible exceptions are the cases excluded in Theorem 1: CS Lagrangian terms having $n=2$ gravitational CS term as the sole gravitational contribution in the case of configurations in which the gauge part of such CS Lagrangian term is not spherically symmetric. When such terms change a solution, then they obviously may change $S_{\mathrm{cov}}$.\footnote{However, as discussed in Sec. \ref{sec:concl}, in such exceptional cases spherical symmetry of the metric will be ruined.}

Let us turn now to the $S_{\mathrm{CS}} = S_{\mathrm{gCS}} + S_{\mathrm{aCS}}$ contribution. Inspection 
of the relevant Lagrangian terms (\ref{mixgCS}) and (\ref{mixaCS}), together with the corresponding entropy formulae (\ref{genwaldent}) and (\ref{Swald}), shows that properties (\ref{rrr0}) imply that all terms give vanishing contribution to the entropy, except those which have just one gravitational factor with $n=2$. These terms have one of the two following forms
\begin{eqnarray}
\CS_2(\Gam) \, \str(\F_1^{r_1}) \cdots \, \str(\F_\NGAUG^{r_\NGAUG})
\label{Th3gCS} \\
\CS_m(\A) \, \str(\F_1^{r_1}) \cdots \, \str(\F_\NGAUG^{r_\NGAUG}) \, \str(\R^2)
\label{Th3aCS}
\end{eqnarray}
The Lagrangian term (\ref{Th3gCS}) produces  a contribution to the entropy formula proportional to 
\cite{BCDPS}
\be \label{Th3gCSent}
\int_{\mathcal{B}} \Gam_N \, \str(\F_1^{r_1}) \cdots \str(\F_\NGAUG^{r_\NGAUG})
\ee
where $(\Gam_N)_\mu = \frac{1}{2} \, \epsilon^\alpha{}_\beta \Gamma^\beta{}_{\alpha\mu}$, while the Lagrangian term (\ref{Th3aCS}) produces a contribution to the  entropy formula proportional to
\be \label{Th3aCSent}
\int_{\mathcal{B}} \CS_m(\A) \, \str(\F_1^{r_1}) \cdots \str(\F_\NGAUG^{r_\NGAUG}) \, \R_N
\ee
where $(\R_N)_{\mu\nu} = \frac{1}{2} \, \tensor{\epsilon}{^\alpha_\beta} \tensor{R}{^\beta_\alpha_\mu_\nu}$. From the fact that in the metric (\ref{staticm}) the components of the binormal $\tensor{\epsilon}{^\alpha_\beta}$ lie in $(t,r)$-plane, and from the explicit 
form of the connection and the Riemann tensor given in (\ref{Gamstat}) and (\ref{Rstat}), it follows that
\be
(\Gam_N)_i = 0 \; , \qquad (\R_N)_{ij} = 0
\ee
This entails that the contributions (\ref{Th3gCSent}) and (\ref{Th3aCSent}) to the black hole entropy
also vanish. This completes the proof of Theorem 3.\footnote{Note that purely gauge CS terms 
(\ref{gaugeCS}), which we included into $\LL_{\mathrm{diff-cov}}$, do not produce additional terms in 
$S_{\mathrm{diff-cov}}$. However, as they contribute to the equation of motion they may change the entropy by affecting the black hole solution.}\\

As a simple example, we apply our results to the special case of CS modified gravity in $D=4$ with the Lagrangian density  
\be \label{4DCSL}
\LL = R \mbox{\boldmath $\epsilon$} + \LL_{\vartheta} + \lambda \CS_2(\Gam) \wedge d\vartheta
\ee
where $\vartheta(x)$ is scalar field, and $\lambda$ is the coupling constant (which does not appear in 
$\LL_{\vartheta}$). As the CS Lagrangian term in these theories can be written in manifestly diff-covariant form $\LL_{\mathrm{CS}} = \vartheta \, \R \wedge \R$, where $\vartheta(x)$ is some scalar field, they do not belong to the type of theories characterized by Lagrangians which cannot be written  in manifestly diff-covariant form, which are of our primary interest. However, our results extend also to these theories and it is interesting to compare them with those existing in the literature on these specific $D=4$ theories.

The theory (\ref{4DCSL}) is rather well-studied\footnote{Mostly in cases $\mathcal{L}_{\vartheta}=0$ (in which case $\vartheta(x)$ is non-dynamical field) and $\mathcal{L}_{\vartheta} = (\partial \vartheta)^2$. See \cite{Alexander:2009tp} for a review.} so we use it to compare our results with the existing literature. Let $g_{0\mu\nu}$ and $\vartheta_0(r,t)$ be some arbitrary spherically symmetric solution of the theory with $\lambda=0$. Then Theorem 1 says that it will be a solution for all values of $\lambda$, in agreement with the known results \cite{Grumiller:2007rv}. Theorem 2 says that, if $\LL_{\vartheta}$ is such that in the theory with $\lambda=0$ Birkhoff's theorem holds, then it holds for all $\lambda$'s. This extends  
\cite{Yunes:2007ss} where such result was shown in case when $\LL_{\vartheta}$ is such that the $\lambda=0$ theory posesses a Schwarzschild black hole as solution. Finally, Theorem 3 says that if this solution describes a black hole, then its entropy does not depend on $\lambda$. In \cite{Grumiller:2008ie} it was shown, using Euclidean methods, that the thermodynamics of such spherically symmetric black holes will not depend on $\lambda$ \emph{modulo} possible contribution of some boundary term 
$\Delta\mathcal{F}$ in the on-shell action which the authors were unable to calculate. Our results show that this unknown boundary term does not influence the entropy of spherically symmetric black holes in theories (\ref{4DCSL}).

\section{Conclusion}
\label{sec:concl}

In this paper we have analyzed the consequences of adding a broad class of Chern-Simons terms to a gravitational action in the case of spherical symmetry. We have considered both gravitational and mixed gauge-gravity CS terms and focused on the case of a general spherically symmetric metric. We have found that in $D>3$ dimensions (the case of the gravitational CS term in 3D must be considered separately and, generally, has already been studied in the literature) the contribution of such terms to the equations of motion vanishes identically, except in the case when the following two conditions are met: (i) a mixed gauge-gravitational CS Lagrangian term, which is a wedge-product of the irreducible gravitational $n=2$ CS term and a purely gauge factor, is present, (ii) this gauge factor (which is by definition gauge invariant) is \emph{not} spherically symmetric for the configuration in question. A consequence is that the gravitational and mixed gauge-gravitational CS Lagrangian terms, apart from the previously mentioned exceptions, do not affect Birkhoff's theorem or any solution with spherically symmetric metric. 

We have then considered the problem of computing the entropy for spherically symmetric black holes in 
the presence of such CS terms. To this end we have used a general formula obtained in a previous paper, 
\cite{BCDPS}, by means of the covariant phase space formalism, adapted to the presence of CS terms. This formula is similar to the one obtained by Wald for covariant gravity Lagrangians. Applied to spherically symmetric black holes it tells us that the contribution of the CS terms to the entropy is null. 

Let us briefly analyze the ``exceptional cases'' in more detail. By Theorem 1, they may appear only for configurations in which the total gauge part of exceptional CS Lagrangian terms is not spherically symmetric. It is important to note that this form is by definition gauge invariant, which means that such exceptional configurations must contain some gauge fields which are not spherically symmetric. Let us now assume that one such configuration is a solution to equations of motion obtained from some Lagrangian 
$\LL_{\mathrm{cov}}$. Then spherical symmetry of the metric requires that the energy-momentum tensor (obtained from $\LL_{\mathrm{cov}}$) be spherically symmetric for such solution. One can now see that ``exceptional'' solutions are somewhat exotic, possessing spherical asymmetry in the gauge fields which disappears in the energy-momentum tensor, but survives in the gauge invariant factor present in the ``exceptional'' CS Lagrangian term we want to add to $\LL_{\mathrm{cov}}$. We are not aware of any explicit examples of such behavior, but we are not aware either of a proof that such cases are not possible in complicated theories with several gauge fields. So, let us now assume that there are such ``exceptional'' solutions, 
and see what we should expect when we add to the Lagrangian some gravitational and mixed gauge-gravitational CS part $\LL_{\mathrm{CS}}$ which includes ``exceptional'' CS terms. This will produce contribution to the equations of motion for the metric, which, for ``exceptional'' terms, is proportional to (in symbolic notation) $\nabla (\R G(\F))$. As $G(\F)$ is for the unperturbed solution not spherically symmetric, we see that in general we should expect a spherically asymmetric perturbation of the metric equation. So, it appears that in general in ``exceptional cases'' one should expect that addition of CS terms completely breaks the spherical symmetry, even for the metric tensor. 

Of course in this paper the hypothesis of spherical symmetry of the metric plays a crucial role, and it is of utmost interest to understand in a systematic way when and how relaxing of this hypothesis will change the null results (in the case of irreducible gravitational CS term in 7-dimensions some specific examples are given in \cite{LuPang}). This is in fact what we intend to investigate next.\\

\vspace{0.5cm}

{\bf Acknowledgements}\\%

\noindent
One of us (L.B.) would like to thank the Theoretical Physics Department, Univ.\ of Zagreb, for hospitality and financial support during his visits there. I.S.\ would like to acknowledge the financial support of CEI Fellowship Programme CERES. Also, M.C., P.D.P., S.P.\ and I.S.\ would like to thank SISSA for hospitality and financial support during visits there and would also like to acknowledge support by the Croatian Ministry of Science, Education and Sport under the contract no.~119-0982930-1016.

\vspace{0.5cm}

\section*{Appendix}
\appendix

\section{Connection and curvature components}

For the general spherically symmetric metric
\be
ds^2 = -f(t,r) \, dt^2 + \frac{dr^2}{g(t,r)} + h(t,r) \, d\Omega^2_{D-2}
\ee
the nonvanishing components of the Christoffel symbols and the Riemann tensor in coordinates 
\refb{gensphcoord} are listed below\footnote{$\dot{f}$ indicates derivative with respect to coordinate $t$, and $f'$ with respect to coordinate $r$},
\be
\Gamma^0_{00} = \frac{\dot{f}}{2f} \qqd \Gamma^0_{11} = -\frac{\dot{g}}{2fg^2} \qqd \Gamma^0_{ii} = \frac{\dot{h}}{2f} \, \Pi(i) \qqd \Gamma^{0}_{01} = \frac{f'}{2f} \qqd \0\\
\Gamma^1_{00} = \frac{g f'}{2} \qqd \Gamma^1_{11} = - \frac{g'}{2g} \qqd \Gamma^1_{10} = -\frac{\dot{g}}{2g} \qqd \Gamma^1_{ii} = -\frac{gh'}{2} \, \Pi(i) \qqd \\
\Gamma^i_{0i} = \frac{\dot{h}}{2h} \qqd \Gamma^i_{1i} = \frac{h'}{2h} \qqd \Gamma^i_{ij} = \ctg \, \theta^j \quad (\textrm{for} \ i > j) \qqd \Gamma^{i}_{jj} = -\ctg \, \theta^i \prod_{k=i}^{j-1} \sin^{2}{\theta^k} \quad (\textrm{for} \ j > i) \0
\ee
\be
R_{0101} = \frac{1}{4} \left( 2f'' - \frac{(f')^2}{f} + \frac{f' g'}{g} - \frac{\dot{f}\dot{g}}{fg^2} - \frac{3(\dot{g})^2}{g^3} + \frac{2\ddot{g}}{g^2} \right) \qqd \0\\
R_{0i0i} = \frac{1}{4} \left( gf'h' + \frac{\dot{f}\dot{h}}{f} + \frac{(\dot{h})^2}{h} - 2\ddot{h} \right) \Pi(i) \qqd \0\\
R_{1i1i} = \frac{1}{4} \left( -\frac{g'h'}{g} - \frac{\dot{g}\dot{h}}{fg^2} + \frac{(h')^2}{h} - 2h'' \right) \Pi(i) \qqd \label{Rspher}\\
R_{0i1i} = \frac{1}{4} \left( -\frac{\dot{g}h'}{g} + \left( \frac{f'}{f} + \frac{h'}{h} \right) \dot{h} - 2\dot{h}' \right) \Pi(i) \qqd \0\\
R_{ijij} = \left( h - \frac{g(h')^2}{4} + \frac{(\dot{h})^2}{4f} \right) \Pi(i) \Pi(j) \0
\ee
In the static case the metric can be written in the form
\be\label{staticm2}
ds^2 = -f(r) dt^2 + \frac{dr^2}{g(r)} + h(r) d\Omega_{D-2}^2
\ee
and the above components reduce to the following nonvanishing ones
\be
\Gamma^{0}_{01} = \frac{f'}{2f} \qqd \Gamma^{1}_{00} = \frac{g f'}{2} \qqd \Gamma^{1}_{11} = - \frac{g'}{2g} \qqd \Gamma^{1}_{ii} = -\frac{gh'}{2} \, \Pi(i) \qqd \Gamma^{i}_{1i} = \frac{h'}{2h} \qqd\0\\
\Gamma^{i}_{ij} = \ctg{\theta^{j}} \quad (\textrm{for} \ i > j) \qqd \Gamma^{i}_{jj} = -\ctg \theta^{i} \prod_{k=i}^{j-1} \sin^{2}{\theta^k} \quad (\textrm{for} \ j > i) \label{Gamstat}
\ee
\be
R_{0101} = \frac{f''}{2} - \frac{(f')^2}{4f} + \frac{f' g'}{4g} &,& \quad R_{0i0i} = \frac{h' f' g}{4} \, \Pi(i) \0\\
R_{1i1i} = \frac{1}{4} \left( -\frac{g'h'}{g} + \frac{(h')^2}{h} - 2h'' \right) \, \Pi(i) &,& \quad 
R_{ijij} = \left( h - \frac{g(h')^2}{4} \right) \Pi(i) \Pi(j) \label{Rstat}
\ee
Due to fact that (\ref{staticm2}) is diagonal, the Riemann tensor components in this case can be written by means of the generalized Kronecker delta symbol,
\be\label{Rsd}
\tensor{R}{^\alpha^\beta_\mu_\nu} = s(\alpha, \beta) \, \delta^{\alpha\beta}_{\mu\nu}
\ee
where $s$ is a symmetric function, $s(x,y) = s(y,x)$, defined implicitly by the components from above.

\section{Important property of spherical Riemann tensor}

Here we shall prove that $\R^3 = 0$ for the general spherically symmetric (not necessary static) metric. The analysis is done case by case. Note that a set equality indicates that the elements are equal up to permutation (e.g. $\{a,b\} = \{0,1\}$ implies that either $a = 0$ and $b = 1$ or $a = 1$ and $b = 0$).
 
$$\left(\R^3\right){}^{\alpha}{}_{\beta\,{\mu_1}\ldots{\mu_6}} = \frac{6!}{2^3} \, \tensor{R}{^\alpha_{\sigma_1}_{[\mu_1}_{\mu_2}} \tensor{R}{^{\sigma_1}_{|\sigma_2|}_{\mu_3}_{\mu_4}} \tensor{R}{^{\sigma_2}_{|\beta|}_{\mu_5}_{\mu_6]}}$$
For $D<6$ this vanishes trivially. For $D \geq 6$, using components of Riemann tensor from the previous Appendix, we see that there are following potentially nonvanishing components of $\R^3$,\\
 
\noindent
1) $\alpha = 0$\\
 
1a) $\sigma_1 = 1$ and $\sigma_2 = 0$ implies that $\{\mu_1,\mu_2\} = \{0,1\} = \{\mu_3,\mu_4\}$;\\
 
1b) $\sigma_1 = 1$ and $\sigma_2 = i$ implies that $i \in \{\mu_1,\mu_2\}$ and $i \in \{\mu_3,\mu_4\}$;\\
 
1c) $\sigma_1 = i$ implies that $i \in \{\mu_1,\mu_2\}$ and $i \in \{\mu_3,\mu_4\}$;\\
 
\noindent
2) $\alpha = 1$ (completely analogous to the first case)\\
 
\noindent
3) $\alpha = i$\\
 
3a) $\sigma_1 = 0$ and $\sigma_2 = 1$ implies $\{\mu_1,\mu_2\} \in \{ \{0,i\} , \{1,i\} \}$ and $\{\mu_3,\mu_4\} = \{0,1\}$;\\
 
3b) $\sigma_1 = 1$ and $\sigma_2 = 0$ implies $\{\mu_1,\mu_2\} \in \{ \{0,i\} , \{1,i\} \}$ and $\{\mu_3,\mu_4\} = \{0,1\}$;\\
 
3c) $\sigma_1 \in \{0,1\}$ and $\sigma_2 = j$ implies $\{\mu_1,\mu_2\} \in \{ \{0,i\} , \{1,i\} \}$, $j \in \{\mu_3,\mu_4\}$ and $j \in \{\mu_5,\mu_6\}$;\\
 
3d) $\sigma_1 = j$ implies $j \in \{\mu_1,\mu_2\}$ and $j \in \{\mu_3,\mu_4\}$;\\
 
\noindent
In all these cases $\R^3$ vanishes identically due to antisymmetrization of indices $\{\mu_1, \dots, \mu_6\}$.\\



\begin{thebibliography}{99}


\bibitem{BCDPS}
  L.~Bonora, M.~Cvitan, P.~Dominis Prester, S.~Pallua and I.~Smoli\'{c},
  {\it Gravitational Chern-Simons Lagrangians and black hole entropy},
  JHEP {\bf 1107} (2011) 085 \
  (arXiv: 1104.2523 [hep-th]).

\bibitem{DJT1}
S.~Deser, R.~Jackiw and S.~Templeton,
{\it Three-dimensional massive gauge theories,}
Phys.~Rev.~Lett. {\bf 48}, 975 (1982)

\bibitem{DJT2}
S.~Deser, R.~Jackiw and S.~Templeton,
{\it Topologically massive gauge theories,}
Ann.~Phys., NY {\bf 140}, 372 (1982)  

\bibitem{Solodukhin:2005ah}
S.~N.~Solodukhin,
{\it Holography with gravitational Chern-Simons,}
Phys.\ Rev.\  {\bf D74} (2006) 024015 \ (arXiv: hep-th/0509148).

\bibitem{Perez:2010hk}
R.~F.~Perez,
{\it Conserved current for the Cotton tensor, black hole entropy and equivariant Pontryagin forms,}
Class.\ Quant.\ Grav.\  {\bf 27} (2010) 135015 \ (arXiv: 1004.3161 [gr-qc]).

\bibitem{Kraus:2005zm}
P.~Kraus, F.~Larsen,
{\it Holographic gravitational anomalies,}
JHEP {\bf 0601} (2006) 022 \ (arXiv: hep-th/0508218).

\bibitem{Park:2006gt}
M.~-I.~Park,
{\it BTZ black hole with gravitational Chern-Simons: Thermodynamics and statistical entropy,}
Phys.\ Rev.\  {\bf D77} (2008) 026011 \ (arXiv: hep-th/0608165).

\bibitem{Miskovic:2009kr}
O.~Mi\v{s}kovi\'{c}, R.~Olea,
{\it Background-independent charges in Topologically Massive Gravity,}
JHEP {\bf 0912 } (2009) 046 \ (arXiv: hep-th/0909.2275).

\bibitem{Alexander:2009tp}
  S.~Alexander and N.~Yunes,
  {\it Chern-Simons Modified General Relativity},
  Phys.\ Rept.\  {\bf 480} (2009) 1 \
  (arXiv: 0907.2562 [hep-th]).

\bibitem{Bertl}
R.~A.~Bertlmann,
{\it Anomalies in quantum field theory},
Oxford University Press, USA (2001).

\bibitem{IyerWald}
V.~Iyer and R.M.~Wald,
{\it Some properties of the Noether charge and a proposal for dynamical black hole entropy,}
Phys.~Rev.~D {\bf 50}, 846 (1994) \ (arXiv: gr-qc/9403028v1).

\bibitem{Tach}
Y.~Tachikawa,
{\it Black hole entropy in the presence of Chern-Simons terms,}
Class.~Quantum Grav. {\bf 24}, 737 (2007) \ (arXiv: hep-th/0611141v2).

\bibitem{Solodukhin:2005ns}
S.~N.~Solodukhin,
{\it Holographic description of gravitational anomalies},
JHEP {\bf 0607} (2006) 003 \ (arXiv: hep-th/0512216).

\bibitem{HawkingEllis}
 S.W.~Hawking and G.F.R.~Ellis,
 {\it The large scale structure of space-time,}
 Cambridge University Press (1973).

\bibitem{Grumiller:2007rv}
  D.~Grumiller and N.~Yunes,
  {\it How do Black Holes Spin in Chern-Simons Modified Gravity?},
  Phys.\ Rev.\  D {\bf 77} (2008) 044015
  (arXiv: 0711.1868 [gr-qc]).

\bibitem{Cantcheff:2008qn}
  M.~B.~Cantcheff,
  {\it Einstein-Cartan formulation of Chern-Simons Lorentz-violating gravity},
  Phys.\ Rev.\  D {\bf 78} (2008) 025002
  (arXiv:0801.0067 [hep-th]).

\bibitem{LuPang}
H.~L\"u and Y.~Pang,
{\it Seven-dimensional gravity with topological terms,}
Phys.~Rev.~D {\bf 81}, 085016 (2010) \ (arXiv: 1001.0042v2 [hep-th]).

\bibitem{Brihaye:2009cc}
  Y.~Brihaye, E.~Radu and D.~H.~Tchrakian,
  {\it AdS(5) solutions in Einstein-Yang-Mills-Chern-Simons theory},
  Phys.\ Rev.\  D {\bf 81} (2010) 064005
  (arXiv: 0911.0153 [hep-th]).

\bibitem{Brihaye:2010wp}
  Y.~Brihaye, E.~Radu and D.~H.~Tchrakian,
  {\it Asymptotically flat, stable black hole solutions in
  Einstein--Yang-Mills--Chern-Simons theory},
  Phys.\ Rev.\ Lett.\  {\bf 106} (2011) 071101
  (arXiv: 1011.1624 [hep-th]).

\bibitem{Brihaye:2011nr}
  Y.~Brihaye, E.~Radu and D.~H.~Tchrakian,
  {\it Einstein-Yang-Mills-Chern-Simons solutions in D=2n+1 dimensions},
  (arXiv: 1104.2830 [hep-th]).

\bibitem{Garcia}
A.A.~Garcia, F.W.~Hehl, C.~Heinicke and A.~Macias,
{\it The Cotton tensor in Riemannian spacetimes,}
Class.~Quantum Grav. {\bf 21}, 1099 (2004) \ (arXiv: gr-qc/0309008v2).

\bibitem{Kaloper:1993kj}
  N.~Kaloper,
  {\it Miens of the three-dimensional black hole},
  Phys.\ Rev.\  D {\bf 48} (1993) 2598
  (arXiv: hep-th/9303007).

\bibitem{Cavaglia}
M.~Cavagli\`a,
{\it The Birkhoff Theorem for Topologically Massive Gravity,}
Grav.\ Cosmol.\ {\bf 5} (1999) 101 \ (arXiv: gr-qc/9904047).

\bibitem{Yunes:2007ss}
  N.~Yunes and C.~F.~Sopuerta,
  {\it Perturbations of Schwarzschild Black Holes in Chern-Simons Modified
  Gravity},
  Phys.\ Rev.\  D {\bf 77} (2008) 064007 \
  (arXiv: 0712.1028 [gr-qc]).

\bibitem{Zegers}
R.~Zegers,
{\it Birkhoff's theorem in Lovelock gravity,}
J.~Math.~Phys. {\bf 46}, 072502 (2005) \ (arXiv: gr-qc/0505016v1).

\bibitem{Deser:2005gr}
  S.~Deser and J.~Franklin,
  {\it Birkhoff for Lovelock redux},
  Class.\ Quant.\ Grav.\  {\bf 22} (2005) L103
  (arXiv: gr-qc/0506014).

\bibitem{Oliva:2011xu}
  J.~Oliva and S.~Ray,
  {\it Birkhoff's Theorem in Higher Derivative Theories of Gravity},
  (arXiv: 1104.1205 [gr-qc]).

\bibitem{Wald1}
R.M.~Wald,
{\it Black hole entropy is the Noether charge,}
Phys.~Rev.~D {\bf 48}, (1993) R3427 \ (arXiv: gr-qc/9307038v1).

\bibitem{JKM}
T.~Jacobson, G.~Kang and R.C.~Myers
{\it On black hole entropy,}
Phys.~Rev.~D {\bf 49}, 6587 (1994) \ (arXiv: gr-qc/9312023v2).

\bibitem{Grumiller:2008ie}
  D.~Grumiller, R.~B.~Mann and R.~McNees,
  {\it Dirichlet boundary value problem for Chern-Simons modified gravity},
  Phys.\ Rev.\  D {\bf 78} (2008) 081502 \
  (arXiv: 0803.1485 [gr-qc]).


\end{thebibliography}
\end{document}